\documentclass{article}
\usepackage[utf8]{inputenc}

\pdfoutput=1
\usepackage{import}
\usepackage{amsmath}
\usepackage{amsthm}
\usepackage{xspace}
\usepackage{comment}
\usepackage{natbib}
\usepackage{url}
\usepackage{hyperref}

\usepackage[capitalise]{cleveref}
\usepackage{tikz}

\usepackage[noend]{algpseudocode}
\usepackage{algorithm}
\usetikzlibrary{decorations.pathreplacing,positioning}

\usepackage{color-edits}
\addauthor[David]{dk}{red}
\addauthor[Yusuf]{yk}{blue}
\addauthor[Vikram]{vk}{green}
\addauthor[Space Saving]{space}{magenta}
\addauthor[Space Saving Yusuf]{spacey}{cyan}

\newtheorem{theorem}{Theorem}[section]
\newtheorem{corollary}[theorem]{Corollary}
\newtheorem{lemma}[theorem]{Lemma}
\newtheorem{proposition}[theorem]{Proposition}
\newtheorem{claim}[theorem]{Claim}
\newtheorem{definition}[theorem]{Definition}

\newcommand{\set}[1]{\left\{ #1 \right\}}
\newcommand{\Set}[2]{\set{#1 \mid #2}}

\newcommand{\floor}[1]{\lfloor {#1} \rfloor}
\newcommand{\ceil}[1]{\left\lceil {#1} \right\rceil}
\renewcommand{\hat}{\widehat}
\renewcommand{\tilde}{\widetilde}

\providecommand{\Kth}[1]{\ensuremath{{#1}^{\rm th}}}

\DeclareMathOperator*{\argmin}{arg\,min}

\def\min{\qopname\relax n{min}}
\def\max{\qopname\relax n{max}}
\def\argmin{\qopname\relax n{argmin}}

\def\X{\mathcal{X}}

\newcommand{\eat}[1]{}

\newcommand{\dsum}{d_{\text{sum}}}
\newcommand{\diam}{\text{diam}}

\providecommand{\ear}{\textsc{Expanding Approvals Rule}\xspace}
\providecommand{\gc}{\textsc{Greedy Capture}\xspace}

\newenvironment{lp*}{\begin{equation*}  \begin{array}{lll}}{\end{array}\end{equation*}}

\usepackage[margin=1in]{geometry}
\title{Proportional Representation in Metric Spaces and Low-Distortion Committee Selection}
\author{Yusuf Hakan Kalayci\thanks{University of Southern California, kalayci@usc.edu} \and David Kempe\thanks{University of Southern California, David.M.Kempe@Gmail.com} \and Vikram Kher\thanks{Yale University, vikram.kher@yale.edu. Work done while at the University of Southern California}}

\begin{document}

\maketitle

\begin{abstract}
    We propose and analyze a natural new definition for when a small set $R$ of $k$ points in a metric space is \emph{representative} of a larger set. 
There is a set $V$ of points to be represented (such as documents or voters), and a set $C$ of candidates (also documents, or candidates for office) who could represent them.
Our definition states (essentially) that for any set $S \subseteq V$ of points comprising a $\theta$ fraction of $V$, the average distance of $S$ to their respective best $\theta k$ points in $R$ should not be larger by more than a factor $\gamma$ compared to their average distance to the best $\theta k$ points among all of $C$.
This definition is a strengthening of the notions of proportional fairness and core fairness, but --- different from those notions --- requires that large cohesive clusters be represented proportionally to their size.

Since there are instances for which --- unless $\gamma$ is polynomially large --- no solutions exist, we study this notion in a resource augmentation framework, implicitly stating the constraints for a set $R$ of size $k$ as though its size were only $k/\alpha$, for $\alpha > 1$.
Furthermore, motivated by the application to elections, we mostly focus on the \emph{ordinal} model, in which the algorithm does not learn the actual distances; instead, the algorithm learns only for each point $v \in V$ and each pair of candidates $c, c'$ which of $c, c'$ is closer to $v$.
Our main result is that the \ear of Aziz and Lee is $(\alpha, \gamma)$ representative in our sense with $\gamma \approx 1 + 6.71 \cdot \frac{\alpha}{\alpha-1}$.

We also obtain three novel byproducts and corollaries from our analysis.
First, we show that the \ear achieves constant proportional fairness in the ordinal model, giving the first positive result on metric proportional fairness with ordinal information.
Second, we show that for the core fairness objective, the \ear achieves the same asymptotic tradeoff between resource augmentation and approximation as the recent results of Li et al., which used full knowledge of the metric.
Finally, our results imply a very simple single-winner voting rule with metric distortion at most 44.

\end{abstract}

\section{Introduction} \label{sec:introduction}

Selecting representatives for a large set is a common and central problem across a wide range of application areas.
As three paradigmatic applications, consider selecting a small set of documents (such as pictures or text) representing a much larger collection, selecting a committee of representatives for a large population, or selecting locations for several public facilities to serve the population of a city. Naturally, there are many ways of defining what it means for a candidate set to be ``representative''; we discuss some key definitions from past work in \cref{sec:related-work}.

A commonly accepted notion of representation is based on \emph{proportionality} \citep{humphreys:proportional-representation,moulin:fair-division}: subgroups of the population should be represented in the selected set proportionally to their size.
Stated differently, if a cohesive subset $S$ comprises a $\theta$ fraction of the documents/population, then (at least) roughly a $\theta$ fraction of the representative set should be similar to the members of $S$.
In terms of documents, this implies that by examining the representative documents, a user can accurately assess the contents of the document collection.
For committee elections, it states that large like-minded groups of the population should be suitably represented in the committee.
And for the location of public facilities, it implies that dense population centers should be sufficiently served with nearby facilities.
Indeed, notions of fairness or representation based on this intuition have been studied extensively, as discussed in \cref{sec:related-work}.

We are particularly interested in the common setting in which the documents or candidates/population are embedded in a metric space in which distances capture dissimilarity.
For documents, this is often the result of feature-based embeddings applied to the documents, and for the selection of public facilities, the metric is naturally derived from geographic proximity or transportation times.
For voters and candidates in elections, the idea of considering all agents as embedded in a metric space, and their preferences being reflective of the distances, was first articulated as \emph{single-peaked preferences} (where the metric space is the line, e.g., representing a one-dimensional left-to-right spectrum of opinions) \citep{black:rationale,moulin:single-peak}, but also subsequently generalized to other metrics \citep{barbera:gul:stacchetti,merrill:grofman}.

Our first main contribution (in \cref{sec:prelims}) is a natural and novel definition of what it means for a set $R$ of $k$ points to ``represent'' a larger set $V$ in a metric space; our notion is a strengthening of the notion of \emph{core fairness} proposed recently by \citet{LLSWW:core-fairness}.

Our second main contribution (in \cref{sec:committee}) is to show that a natural algorithm (a special case of the \ear of \citet{aziz:lee:expanding-approvals-rule}) achieves strong representativeness guarantees for the new definition.
It does so even though it works in the \emph{ordinal} model (see \cref{sec:ordinal}), in which the algorithm only learns, for each voter/document/citizen, the ranking of potential representatives by increasing distances (but not the distances themselves).
As immediate corollaries of our analysis, we obtain the first algorithm with constant proportional fairness for metric costs in the ordinal model, an algorithm with ordinal information achieving --- up to constants --- the same parameter tradeoff for approximate core fairness as the one of \citet{LLSWW:core-fairness} (which had access to the full metric), as well as an extremely simple single-winner voting rule with constant metric distortion; see \cref{sec:committee}.

Finally, in \cref{sec:known-distances}, we show that when the algorithm has access to all the distances, a slight modification of the \gc algorithm of \citet{pfc} provides improved constants in the representativeness guarantees.

\section{The Key Fairness Concepts} \label{sec:prelims}

Our first main contribution is a new definition of a representative set in a metric space.
Our definition is a strengthening of the notion of core fairness proposed by \citet{LLSWW:core-fairness}, and --- as that definition does --- naturally recovers an approximate median as a special case for $k=1$.

We consider settings in which a set $V$ should be ``well represented'' by a subset of a set\footnote{$V=C$ --- the most natural case when selecting documents, and a case corresponding to peer selection in the context of elections --- is of course allowed.} $C$; for examples, see \cref{sec:introduction}.
We write $n=|V|$ and $m=|C|$ for the sizes of the sets, and $k < m$ for the size of the subset of $C$ that is to be selected.
For concreteness in our nomenclature, we will refer to $V$ as \emph{voters} and $C$ as \emph{candidates} throughout, although we do not exploit any specific properties of this domain. 

We assume that $V \cup C$ is embedded in a (pseudo-)metric space $(V \cup C, d)$.\footnote{Recall that a \emph{metric} is a non-negative function $d$ on pairs satisfying that $d(x,x) = 0$ for all $x$, \emph{symmetry} ($d(x,y) = d(y,x)$ for all $x, y$), triangle inequality ($d(x,z) \leq d(x,y) + d(y,z)$ for all $x, y, z$), and \emph{positivity} ($d(x,y) > 0$ whenever $x \neq y$). A pseudo-metric is allowed to violate positivity, i.e., multiple points of the metric space can be at distance 0 from each other.}
The goal is to pick a set $R \subseteq C$ of $k$ \emph{representatives} to ensure that each sufficiently large subset $S \subseteq V$ is ``well represented'', in a sense we define next.
In keeping with the voting-related nomenclature, we will refer to $R$ as a \emph{committee} and to $S$ as a \emph{coalition}.

We write $\dsum(X,Y)=\sum_{x\in X, y \in Y} d(x,y)$ for the sum of distances between all pairs in $X \times Y$.
When $X = \set{x}$ is a singleton, we write $\dsum(x, Y) = \dsum(\set{x}, Y)$.

\subsection{Proportional Representation} \label{sec:keydef}

Recall that our goal is to ensure that all sufficiently large coalitions of voters are well represented.
Specifically, if a coalition $S$ comprises a $\theta$ fraction of all \emph{voters}, at least a $\theta$ fraction of the \emph{committee} should be approximately closest to $S$ as a whole.
To phrase this requirement cleanly, we recall the definition of the \emph{Hare Quota} $p=\ceil{n/k}$, the number of voters/documents/citizens represented by any one selected candidate/document/facility.
As articulated by \citet{LLSWW:core-fairness} in their definition of (approximate) \emph{core fairness}, any coalition $S$ whose size is at least the Hare Quota should have at least one roughly satisfactory representative in $R$.
Our generalization requires that, for any positive integer $t$, each member of a coalition $S$ of size $|S| \geq t \cdot p$ should have at least $t$ representatives in $R$, such that no other $t$ candidates are much better for the coalition compared to their individual best $t$ members in $R$.
We now formally define what it means for the subset of the committee to be ``approximately optimal'' for all coalitions.

\begin{definition}[$\gamma$-proportionally representative committee]
  \label{def:gamma-proportional}
  A committee $R$ is called \emph{$\gamma$-proportionally representative} if for every coalition $S \subseteq V$ of size at least $t \cdot p$, the committee satisfies:
  \begin{align}
    \sum_{v \in S} \min_{\substack{R'_{v} \subseteq R\\|R'_{v}|=t}} \dsum(v, R'_{v})
  & \leq \gamma \cdot
    \min_{\substack{C' \subseteq C\\|C'|=t}} \dsum(S, C').
    \label{eqn:gamma-proportional}
  \end{align}
\end{definition}
  
\cref{def:gamma-proportional} captures a notion of approximate \emph{stability}: no sufficienty large coalition $S$ of voters could find an alternative committee for themselves of corresponding size $t$ which they strongly prefer over their individually best size-$t$ subcommittees of $R$.
Note that proportional representation (\cref{def:gamma-proportional}) is a more demanding requirement than approximate core fairness in the sense of \citet{LLSWW:core-fairness}, as defined in \cref{def:alpha-beta-core}; we will elaborate on this in more detail below.
A positive feature of \cref{def:gamma-proportional} is that it does not require a notion of ``cohesive'' coalitions which must be represented.
Rather, a requirement for \emph{all} sufficiently large coalitions is given; however, the requirements for very spread-out coalitions are typically trivially satisfied, because such coalitions do not have attractive alternatives to deviate to. 
  
While natural and intuitive, our definition of proportional representation
is unfortunately too demanding; \citet{LLSWW:core-fairness} already showed that there are instances for which the $(1,\Omega(\sqrt{n}))$-core\footnote{The instances described in \citet{LLSWW:core-fairness} require $n=\Theta(k^2)$. In \cref{ex:resource-augmentation}, we give a simple class of instances showing a slightly more fine-grained lower bound of $\Omega(\min(k,n/k))$, thus giving a lower bound for the entire range of $k$.} is empty. Since \cref{def:gamma-proportional} has additional constraints, we can in general not hope for $\gamma = o(\sqrt{n})$.

Therefore, as in \citet{LLSWW:core-fairness}, we relax \cref{def:gamma-proportional} by allowing for \emph{resource augmentation} --- see \citet{jiang-approximately-stable-committee} for another example and discussion of the use of resource augmentation to deal with impossibility of proportional representation.
Specifically, for a \emph{resource augmentation parameter} $\alpha \geq 1$, we consider the problem of selecting a committee of size $k$, but require the weaker stability guarantee for a committee of (smaller) size $k/\alpha$, as captured by the following definition.

\begin{definition}[$(\alpha, \gamma)$-proportionally representative committee]
  \label{def:alpha-gamma-proportional}  
  For a resource augmentation parameter $\alpha \geq 1$, a committee $R$ is called \emph{$(\alpha,\gamma)$-proportionally representative} if for every coalition $S \subseteq V$ of size at least $t \cdot \alpha \cdot p$, the committee satisfies: 
  \begin{align}
    \sum_{v \in S} \min_{\substack{R'_{v} \subseteq R\\|R'_{v}|=t}} \dsum(v, R'_{v})
 & \leq \gamma \cdot \min_{\substack{C' \subseteq C\\|C'|=t}} \dsum(S, C').
  \label{eqn:alpha-gamma-proportional}
  \end{align}
\end{definition}

\cref{def:alpha-gamma-proportional} is a strengthening of the definition of approximate core in \citet{LLSWW:core-fairness}, defined as follows:

\begin{definition}[Approximate Core]
  \label{def:alpha-beta-core}  
  For a resource augmentation parameter $\alpha \geq 1$, a committee $R$ is in the \emph{$(\alpha,\beta)$-core} if for every coalition $S \subseteq V$ of size at least $\alpha \cdot p$, the committee $R$ satisfies: 
  \begin{align}
    \sum_{v \in S} \min_{r_v \in R} d(v,r_v)
 & \leq \beta \cdot \min_{c \in C} \dsum(S, c).
  \label{eqn:alpha-beta-core}
  \end{align}
  
\end{definition}  

Thus, the definition of the $(\alpha,\beta)$-core\footnote{We use $\gamma$ in place of $\beta$ in \citet{LLSWW:core-fairness} to emphasize the difference in the definitions.} is obtained by requiring only the constraints for $t=1$ to hold in \cref{def:alpha-gamma-proportional}.

After augmenting resources, the definition becomes less stringent, and one may naturally inquire if approximation in the objective can be avoided, i.e., whether $\gamma=1$ (or $\beta=1$) can be achieved.
We show that this is in general not possible; in \cref{ex:away-from-one}, we prove the following lower bound, which already holds for the weaker notion of approximate core in the sense of \citet{LLSWW:core-fairness}:

\begin{proposition} \label{prop:bounded-from-optimal}
  For every $\alpha \geq 1$, there are instances for which the $(\alpha, 1+1/(2\alpha))$-core is empty, and so, no $(\alpha, 1+1/(2\alpha))$-proportional representation exists.
\end{proposition}

Furthermore, in \cref{ex:info-hardness}, we show that when $\alpha \to 1$, a large blowup in $\gamma$ is unavoidable in general. 
Again, we show this lower bound result already for the approximate core:

\begin{proposition} \label{prop:diverging}
  There are $\alpha$ arbitrarily close to 1 and corresponding instances for which the $(\alpha, 1/(4(\alpha-1)))$-core is empty, and so, no $(\alpha, 1/(4(\alpha-1)))$-proportional representation exists.
\end{proposition}
Note that \cref{prop:diverging} provides an asymptotically matching lower bound for Theorem~19 of \citet{LLSWW:core-fairness}.

While we cannot achieve $\alpha \approx 1$ or $\gamma \approx 1$ without a large blowup in the other parameter, we still seek to design (polynomial-time) algorithms for computing committees achieving a good tradeoff between $\alpha$ and $\gamma$; indeed, this is the main goal of our work, studied in \cref{sec:committee}.

\subsection{Proportional Representation, Core Fairness, and Proportionally Fair Clustering}
\label{sec:pr-vs-pfc}

As discussed above, proportional representation is a natural strengthening of the notion of core fairness.
Another closely related concept is \emph{proportional fairness}, introduced in the field of clustering by \citet{pfc}, and studied further by \citet{pfc-revisited}.
Here, the setup is the same, and the committee $R$ is construed as \emph{cluster centers}:

\begin{definition}[$\gamma$-proportional fairness]
  \label{def:proportional-fairness}
  A committee $R$ of size $k$ is \emph{$\gamma$-proportionally fair} if for every voter coalition $S$ of size at least $p$ and every alternate candidate $c$, at least one voter $v \in S$ satisfies $\min_{r \in R} d(v, r) \leq \gamma \cdot d(v, c)$.
\end{definition}

\citet{LLSWW:core-fairness} already pointed out that every committee in the $(1,\beta)$-core according to their definition is also $\beta$-proportionally fair. Since $(1,\beta)$-proportional representation implies membership in the $(1,\beta)$-core, it is strictly stronger than $\beta$-proportional fairness.

We discuss key differences between \cref{def:gamma-proportional,def:alpha-beta-core,def:proportional-fairness}, as well as their implications.
We believe that they justify \cref{def:gamma-proportional} as a more suitable notion of representativeness of a committee.

\begin{enumerate}
\item Regardless of the size of $S$ and $k$, the definition of approximate core (and thus also proportionally fair clustering) only requires the existence of a single candidate $c$ that is preferred in order to ``satisfy'' $S$.
As a result, a large committee $R$ could significantly distort the composition of $V$. 

For example, consider $k=100$, with two clusters $V_1, V_2$, containing $99\%$ and $1\%$ of the voters, respectively. 
The two clusters are far from each other, and each has a large number of possible candidates, all at distance 1 from each voter in the cluster. 
Then,
$R$ could contain 99 candidates close to $V_2$ and one candidate close to $V_1$, while satisfying \cref{def:alpha-beta-core} and \cref{def:proportional-fairness} with $\beta = 1$.
The exact same types of distances can be used to obtain similarly non-proportional outcomes for Example~1 in \citet{LLSWW:core-fairness}.
In a sense, neither \cref{def:alpha-beta-core} nor \cref{def:proportional-fairness} include a notion of \emph{proportionality}, and thus, they fail to achieve it.

\item \cref{def:proportional-fairness} suffers from a second weakness, which is fixed by \cref{def:alpha-beta-core}: for small $k$ (in particular, $k=1$), the definition is extremely lenient. In fact, for $k=1$, it allows the chosen candidate to be any candidate not Pareto-dominated by another; among others, this includes any candidate ranked first by at least one voter.
  This is a very weak requirement for a chosen candidate being ``representative''.
  In contrast, any candidate in the $(1,\beta)$-core (and thus any candidate who is $(1,\beta)$-representative) must be a $\beta$-approximate median of the voters, a much more meaningful sense of being representative.

\item The distinction between the definitions can also be viewed through the lens of whether the utility from deviations is transferable within the deviating coalition, as pointed out by \citet{LLSWW:core-fairness}. 
  While proportionally fair clustering implicitly assumes non-transferable utilities, \cref{def:alpha-beta-core} as well as \cref{def:gamma-proportional} correspond to transferable utilities, leading to a larger set of ``deviation threats''.
  Notice that the question of whether utility is transferable also arises in more ``classical'' definitions of the core (e.g., \citet{peters-core-transferable-utility}).
\end{enumerate}

\section{The Ordinal Information Model} \label{sec:ordinal}

We have so far stated the problem of finding an $(\alpha,\gamma)$-proportionally representative committee in a model where the metric $d$ is known to the algorithm.
This model is a natural fit for the applications of selecting representative documents or locations for facilities: the documents are typically embedded into a (often Euclidean) metric space by a known algorithm, so all distances are known; similarly, the metric space for the citizens and facilities is defined by known geography or travel times.

In contrast, for the election of a committee, parliament, or similar body, the metric space is a useful modeling tool, but is typically not known explicitly.
Rather, the voters are assumed to \emph{rank} the candidates by non-decreasing distance from themselves, and the algorithm has access to the rankings, but not the distances.
Despite this limited information (after all, many different metrics may be consistent with the rankings),
an algorithm or voting rule should select an approximately optimal solution, in our case, a proportionally representative set $R$ exhibiting a good tradeoff between $\alpha$ and $\gamma$.

This framework is called the \emph{ordinal information model} of the \emph{metric distortion}\footnote{A parallel line of work (e.g., \citep{BCHLPS:utilitarian:distortion,boutilier:rosenschein:incomplete,procaccia:rosenschein:distortion}) considers the same tradeoff of ordinal vs.~cardinal information in a setting in which the rankings are derived from positive \emph{utilities} rather than distances/costs.} framework \citep{anshelevich:bhardwaj:elkind:postl:skowron,anshelevich:filos-ratsikas:shah:voudouris:reading-list,anshelevich:filos-ratsikas:shah:voudouris:retrospective,metric-multi}, contrasting it with the \emph{cardinal} model, in which the metric is known explicitly.
The worst-case loss in the objective function due to the lack of information is called \emph{(metric) distortion}, and has been studied for numerous optimization problems.
Most notable is the by now extensive line of work on the distortion of \emph{single-winner elections} \citep{anshelevich:bhardwaj:postl,anshelevich:bhardwaj:elkind:postl:skowron,munagala:wang:improved,gkatzelis:halpern:shah:resolving,PluralityVeto,gkatzelis:latifian:shah:best-of-both}; however, more complex objectives have also been studied \citep{anshelevich:filos-ratsikas:shah:voudouris:retrospective,anshelevich:filos-ratsikas:shah:voudouris:reading-list,anshelevich:zhu:ordinal-approximation,anari:charikar:ramakrishnan:distortion-metric-matching}.

Viewed in this context, under the ordinal information model, our goal of selecting a committee $R$ can be viewed as a natural extension of the metric distortion objective to multi-winner elections; a detailed comparison to other recently proposed multi-winner distortion objectives is given in \cref{sec:related-work}.

We now define the concepts formally.
The algorithm learns, for each voter $v \in V$, a \emph{ranking} of candidates $\succ_v$.
Voters rank candidates by non-decreasing distances, so $c \succ_v c'$ implies that $d(v,c) \leq d(v,c')$.
We use $\pi_v(c)$ to denote the position of candidate $c$ in $v$'s ranking, with $\pi_v^{-1}(1)$ being $v$'s most preferred candidate.
We write $\succ_V = (\succ_v)_{v \in V}$ for the vector of all voters' rankings, and refer to it as the \emph{ranked-choice profile}.
An \emph{election} consists of the triple $(V, C, \succ_V)$.
We say that a pseudo-metric $d$ is \emph{consistent} with the ranked-choice profile $\succ_V$ if it satisfies that $d(v,c) \leq d(v,c')$ whenever $c \succ_v c'$, for all $v, c, c'$.

An ordinal committee selection rule $f$ receives as input the committee size $k$ and the election $(V, C, \succ_V)$, and outputs a committee $R \subseteq C$ of size $k$.

Our notion of proportional representation for ordinal models in committee selection rules is closely related to, and intended to be a natural generalization of, the concept of metric distortion for single-winner elections.
Recall that the \emph{metric distortion} of a single-winner voting rule $f$ is the worst-case ratio (over all elections, and all metrics consistent with the election) of the total cost of the chosen winner relative to the total cost of the optimum candidate, i.e.,
  \[
  \max_{(V,C,\succ_V)} \max_{d \text{ consistent with } \succ_V} \frac{\dsum(V,f(V,C,\succ_V))}{\min_{c \in C} \dsum(V,c)}.
  \]
As a result, notice that in the special case when $k=1$ and $\alpha=1$, a committee selection rule $f$ is $\gamma$-proportionally representative if and only if $f$ is a single-winner voting rule with metric distortion at most $\gamma$.
This is because the only coalition of size $p = n$ is $S = V$, so
\cref{eqn:alpha-gamma-proportional} requires that for the chosen candidate $R = \set{\hat{c}} = f(V,C,\succ_V,k)$, we have $\dsum (V, \hat{c}) \leq \gamma \cdot \dsum (V, c)$ for every candidate $c$; in particular, the optimal candidate $c = c^*$.

\section{Our Main Result} \label{sec:committee}

Our second main contribution --- and the key technical work in this paper --- is to show that a very natural algorithm, namely, a special case of the \ear of \citet{aziz:lee:expanding-approvals-rule}, selects an $(\alpha,\gamma)$-proportionally representative committee $R$ of size $k$ which achieves constant $\alpha$ and $\gamma$; furthermore, it does so even in the ordinal model.

The algorithm runs in iterations, parametrized by a \emph{tolerance} parameter $\tau$, which starts at 1 and increases in each round.
In iteration $\tau$, the algorithm considers, in some arbitrary order, each (remaining) voter as approving their top $\tau$ choices.
As soon as at least $p=\ceil{n/k}$ of the remaining voters approve of a particular candidate $c$, this candidate is added to the committee, and those $p$ voters are permanently removed from further consideration. 
We will say that $c$ \emph{covers} these voters.
The fact that the algorithm processes the voters in an arbitrary order (instead of simultaneously) in each iteration achieves an (arbitrary) tie breaking implicitly, so the algorithm does not have to consider ties between multiple candidates becoming eligible for inclusion.
This process continues until a committee of size $k$ has been formed; if the algorithm would otherwise terminate with fewer candidates, it adds arbitrary candidates to ensure that the committee has size $k$. The algorithm is described formally as \cref{alg:URCS}.

\begin{algorithm}
    \caption{\ear} \label{alg:URCS}
    
    \textbf{Input: } Election $(V, C, \succ_V)$, Committee Size $k$
    
    \textbf{Output: } Committee $R$
    
    \begin{algorithmic} 
        \State Let $U \leftarrow V$ be the set of uncovered voters.
        \State Let $R \leftarrow \emptyset$ be the selected committee.
        \State Let $N_c \leftarrow \emptyset$ for all $c \in C$.
        \For{$\tau = 1, \ldots, m$}
           \For{$v \in V$ in arbitrary order}
              \If{$v \in U$}
                 \State Let $c = \pi_v^{-1}(\tau)$.
                 \If{$c \notin R$}
                    \State Let $N_c \leftarrow N_c \cup \set{v}$, i.e., add $v$ to the neighborhood of $c$.
                    \If{$|N_c|=\ceil{n/k}$}
                        \State $R \leftarrow R \cup \set{c}$, i.e., include $c$ in the committee.
                        \State $N_{c'} \leftarrow N_{c'} \setminus N_c$ for all $c' \in C\setminus R$, i.e., update the neighborhood for non-committee candidates.
                        \State $U \leftarrow U \setminus N_c$, update the set of uncovered voters.
                        \State We say that $N_c$ has been \emph{covered} by $c$.
                     \EndIf
                  \EndIf
               \EndIf
            \EndFor
            \EndFor
            \State \textbf{if} $|R| < k$ \textbf{then} add $k - |R|$ arbitrary candidates to $R$.
    \end{algorithmic}
\end{algorithm}

\begin{theorem} \label{thm:main}
     The \ear outputs a committee $R$ of size $k$ which is $(\alpha, \gamma(\alpha))$-proportionally representative for all $\alpha > 1$, with 
      $\gamma(\alpha) = {1 + \frac{7+\sqrt{41}}{2}\cdot \frac{\alpha}{\alpha-1} \approx 1 + 6.71 \cdot \frac{\alpha}{\alpha-1}}$.
\end{theorem}

\cref{thm:main} gives guarantees only for $\alpha > 1$, and \cref{prop:diverging} shows that this is unavoidable for decent approximation guarantees, unless $k$ is very small or large.
When resources are not augmented, i.e., for $\alpha=1$, we obtain the following weaker guarantee, which is proved in \cref{sec:without_augmentation}.

\begin{theorem} \label{thm:no-augmentation}
  The output $R$ of \cref{alg:URCS} is $(1, O(n/k))$-proportionally representative.
\end{theorem}

Our analysis directly implies several novel results.
First, an immediate corollary of a key lemma in the proof of \cref{thm:main} is that the \ear achieves proportional fairness 5.71; this constitutes the first result achieving constant proportional fairness with metric costs in the ordinal information model:

\begin{corollary} \label{cor:pfc}
    The \ear is a $\frac{5+\sqrt{41}}{2} \approx 5.71$-proportionally fair clustering algorithm under the ordinal information model with metric costs.
\end{corollary}

The constant $\frac{5+\sqrt{41}}{2}$ is larger than the best proportional fairness achievable with full knowledge of the metric space, which is $1+\sqrt{2} \approx 2.41$ \citep{pfc,pfc-revisited}.
This is perhaps not surprising, given that in the ordinal information model, the algorithm is missing crucial information.
Indeed, \cref{ex:ordinal-vs-cardinal} gives an instance in which under the ordinal information model, no deterministic algorithm can produce a $\gamma$-proportionally fair committee for any $\gamma < 2+\sqrt{5} \approx 4.23$; thus, there is necessarily a gap between the best proportional fairness achievable in the cardinal and ordinal models.
Obtaining the best possible proportional fairness guarantees under metric costs and ordinal information is an interesting direction for future work.

A second corollary can be immediately obtained from \cref{thm:main} and \cref{thm:no-augmentation}: because $(\alpha,\beta)$-representativeness implies being in the $(\alpha,\beta)$-core, the guarantees of \cref{thm:main} and \cref{thm:no-augmentation} apply verbatim to the latter. Thus, the \ear achieves approximate core fairness in the sense of \citet{LLSWW:core-fairness} in the ordinal metric cost model; this, too, is the first positive result on core fairness in the ordinal metric cost model.

\begin{corollary} \label{cor:core-fairness}
  The committee $R$ output by \cref{alg:URCS} is in the $(\alpha, \beta(\alpha))$-core for all $\alpha > 1$, with $\beta(\alpha) = {1 + \frac{7+\sqrt{41}}{2} \cdot \frac{\alpha}{\alpha-1} \approx 1 + 6.71 \cdot \frac{\alpha}{\alpha-1}}$.
  It is also in the $(1, O(n/k))$-core.
\end{corollary}

Note that these bounds match the information-theoretic lower bound of \cref{prop:diverging} and the lower bound of $\Omega(\min(k,n/k))$ for $\alpha=1$ in \cref{ex:info-hardness} up to constant factors.
They mirror (with slightly worse constants) the bounds obtained by \citet{LLSWW:core-fairness} in the model with known distances (Theorem~19).

A third corollary, also from \cref{thm:main}, gives an extremely simple single-winner voting rule with constant metric distortion.

\begin{corollary} \label{cor:single-winner-distortion}
  For a given set of candidates $C$ and voters $V$, consider the following voting rule: find a candidate $c$ who is in the top $\tau$ positions of at least $\ceil{n/2}$ voters, for the smallest possible $\tau$. (Break ties arbitrarily.) Find a candidate $c'$ who is in the top $\tau'$ positions of the remaining $\floor{n/2}$ voters, for the smallest possible $\tau'$. (Again, break ties arbitrarily.) Return the one of $c,c'$ preferred by a majority of voters.

  This voting rule has metric distortion at most $44$ for the single-winner election $(V,C, \succ_V)$.
\end{corollary}

While the distortion guarantee of $44$ given by \cref{cor:single-winner-distortion} is worse than the (optimal) metric distortion of $3$ achieved by
\citep{gkatzelis:halpern:shah:resolving,PluralityVeto}, this voting rule is arguably even simpler than the rules previously known to achieve constant metric distortion
(\textsc{Copeland} \citep{anshelevich:bhardwaj:elkind:postl:skowron}, \textsc{Plurality-Matching} \citep{gkatzelis:halpern:shah:resolving}, \textsc{Plurality-Veto} \citep{PluralityVeto}).

\subsection{Running Time Analysis and Discussion}

We briefly analyze the running time of \cref{alg:URCS}, and explain why it is a special case of the \ear, as well as its relationship to another recently proposed algorithm.

We begin with a brief running time analysis.
The number of basic set operations (addition or removal of an element to/from a neighborhood, or test of membership in a set) is $O(mn)$. This is because each voter is added and removed to/from each neighborhood at most once, and at any time, each neighborhood contains only uncovered voters.\footnote{Notice that the two nested loops with an internal update for all non-committee candidates' neighborhoods initially suggest $O(m^2n)$. By keeping track of the set of all neighborhoods $N_c$ that a given voter $v$ has been included in, the updating can be implemented to only incur time for $(v,c)$ pairs such that $v \in N_c$ at the time.}
Set operations can be implemented in time $O(\log n)$ or $O(\log m)$, giving a running time of $\tilde{O}(nm)$.
Notice that because specifying all rankings requires space $\Omega(n m \log m)$, the running time is essentially linear.\footnote{A slight exception is when the number of voters $n$ is superpolynomially large in $m$, in which case operations for the sets $N_c$ may lead to slightly larger running time. 
}
  
We next discuss why \cref{alg:URCS} is a special case of the \ear.
The \ear, as presented by \citet{aziz:lee:expanding-approvals-rule}, follows the same pattern of increasing $\tau$, and including a candidate when $N_c$ (in our notation) is large enough.
Because \citet{aziz:lee:expanding-approvals-rule} expand all $N_c$ simultaneously, multiple candidates may become eligible for inclusion at the same time.
In this case, \citet{aziz:lee:expanding-approvals-rule} use a fixed order on candidates to break ties.
Since the order in which \cref{alg:URCS} processes candidates in each iteration is arbitrary (and can differ from iteration to iteration), one could choose the order to emulate the ordering that would be chosen in \citet{aziz:lee:expanding-approvals-rule}.
A second consequence of the simultaneous updating of the $N_c$ in \citet{aziz:lee:expanding-approvals-rule} is that $|N_c| > \ceil{n/k}$ becomes possible.
\citet{aziz:lee:expanding-approvals-rule} have all voters start with weights $w_v = 1$, and reduce the combined weights of voters in $N_c$ by a total\footnote{In fact, \citet{aziz:lee:expanding-approvals-rule} also allow for slight variations in the total amount of weight.} of $\ceil{n/k}$; this includes reducing the weights of voters fractionally, or selecting an arbitrary subset of size $\ceil{n/k}$ who are eliminated by virtue of having their weights reduced to 0.
We believe that our proportional representation results hold for the \ear in full generality; however, various steps of the analysis are cleaner to formulate and argue using our approach of avoiding ties and fractional weights.

The \ear is also somewhat similar to Algorithm~C of \citet{skowron-proportional-representation-utilitarian}.
The work of \citet{skowron-proportional-representation-utilitarian} is concerned with selecting committees under the rules of \citet{monroe-fully-proportional-representation} and \citet{chamberlin-courant}.
  Both of these rules define pairwise dissatisfaction values between voters and candidates, increasing in how far down in the voter's ranking the candidate appears.
  The goal is to select a committee minimizing the sum of dissatisfactions of all voters, where the dissatisfaction of a voter with a committee is their dissatisfaction with their favorite committee member.
  Algorithm~C of \citet{skowron-proportional-representation-utilitarian} also greedily selects candidates one by one, always choosing next the candidate reducing dissatisfaction the most.
  Notice, however, that this is not identical to expanding approvals, as voters may see their dissatisfaction improve multiple times, and there is no hard ``inclusion threshold'' akin to our value of $\tau$.

\subsection{Stronger Proportional Fairness: The Key Lemma}
\label{sec:missing-proof-stability}

  In the analysis of \cref{alg:URCS}, a central concept is for any coalition $S$ the set of all chosen representative candidates $r \in R$ who covered at least one voter in $S$.
We formally define this notion; recall here that $N_r$ is defined in \cref{alg:URCS}:
  
\begin{definition}[Representatives for a Coalition] \label{def:coalition-representatives}
The set of representatives for the coalition $S \subseteq V$ is defined as $R[S] = \Set{r \in R}{N_r \cap S \neq \emptyset}$, i.e., $R[S] \subseteq R$ is the set of candidates $r \in R$ whose neighborhood contains at least one voter in $S$.
\end{definition}

We show that the committee $R$ returned by \cref{alg:URCS} satisfies a somewhat stronger notion of proportional fairness, i.e., a modification of \cref{def:proportional-fairness}.
\cref{lem:stability} shows that for any coalition $S$, the representatives $R[S]$ are already sufficiently attractive that $S$ will not unanimously deviate.
The guarantee differs from \cref{def:proportional-fairness} only in the slight strengthening of replacing $R$ with $R[S]$.

\begin{lemma} \label{lem:stability}
  The committee $R$ output by \cref{alg:URCS} has the following stability property, with $\rho = \frac{5+\sqrt{41}}{2}\approx 5.71$:
  For every coalition $S$ of size $|S| \geq p = \ceil{n/k}$, there exists a voter $v \in S$ with

  \begin{align*}
    \min_{r \in R[S]} d(v,r)
    & \leq \rho \cdot \min_{c \in C \setminus R} d(v,c).
  \end{align*}

\end{lemma}

Because $R[S] \subseteq R$, \cref{lem:stability} immediately implies that \cref{alg:URCS} outputs a $\frac{5+\sqrt{41}}{2}$-proportionally fair clustering in the ordinal information model, which proves \cref{cor:pfc}.

We now prove \cref{lem:stability}.
As in the work of \citet{pfc-revisited}, our analysis is based on the notion of Apollonius's Circle, defined as follows:

\begin{definition}[Apollonius's Circle] \label{ref:apollonius}
    Let $(\X, d)$ be an arbitrary metric space, and $x,y \in \X$ arbitrary points of the space. For any $\rho > 1$, let $A(x,y,\rho) = \Set{z \in \X}{d(x,z) \cdot \rho \leq d(y,z)}$.

  $A(x,y,\rho)$ is called the \emph{Apollonius's Circle} of $x$ and $y$ with distance ratio $\rho$.
\end{definition}

We are specifically interested in the diameter\footnote{Recall that the diameter of a (closed) set $S$ of points is defined as $\diam(S) = \max_{x,y \in S} d(x,y)$.} of Apollonius's Circles.
The following lemma due to \citet{pfc-revisited} states an upper bound on the diameter of Apollonius's Circles in arbitrary metric spaces.

\begin{lemma}[Theorem~4 of \citet{pfc-revisited}] \label{lem:apollonius-diam}
  For any metric space $(\X,d)$, points $x,y \in \X$ and constant $\rho > 1$, the Apollonius's Circle $A(x,y,\rho)$ has diameter at most
        \begin{align*}
            \diam(A(x,y,\rho)) & \leq \frac{2}{\rho-1} \cdot d(x,y).
        \end{align*}
\end{lemma}

\begin{proof}[Proof of Lemma~\ref{lem:stability}]
  Consider a coalition $S$ of size $|S| \geq p$ and let $c^* \in C \setminus R$ be a candidate satisfying the condition
  \begin{align}
    \rho \cdot d(v,c^*) & \leq \min_{r \in R[S]} d(v,r)
                          \qquad \text{ for all } v \in S.
          \label{eqn:other-candidate}
  \end{align}
We will show that under the assumption \eqref{eqn:other-candidate}, $\rho$ must satisfy $\rho \leq \frac{5+\sqrt{41}}{2}$.

We first show that $R[S]$ is non-empty. 
To do so, note first that in the last iteration (i.e., for $\tau = m$), for all remaining candidates $c$, the neighborhood is $N_c = U$. 
The set of remaining candidates is also non-empty, because $c^* \notin R$.
If $|U| \geq p$ at termination, the last iteration of \cref{alg:URCS} would have added at least one more candidate.
Therefore, at termination, the set of uncovered voters has size $|U| < p \leq |S|$.
As a result, at least one voter in $S$ must have been covered by \cref{alg:URCS}, meaning that $R[S] \neq \emptyset$.

Let $r^* \in R[S]$ be the first candidate added to the committee with $N_{r^*} \cap S \neq \emptyset$, and $\tau$ the tolerance level of the algorithm at the time of addition.
Let $v \in S \cap N_{r^*}$ be an arbitrary voter in $S$ who was among the first to be covered by \cref{alg:URCS}.
We show that $v$ cannot be too much further from $r^*$ than from $c^*$.

\begin{claim} \label{cl:distance-diameter}
  $d(v,r^*) \leq 2 \cdot \diam(S \cup \set{c^*})$.
\end{claim}
\begin{proof}
  Let $R_{< \tau} \subseteq R$ be the set of all candidates included by \cref{alg:URCS} for tolerance values strictly smaller than $\tau$, and $\hat{C} = C \setminus R_{< \tau}$.
  By definition of $r^*$ and $\tau$, we have that $R[S] \subseteq \hat{C}$ and $c^* \in \hat{C}$.

  Suppose for contradiction that $d(v,r^*) > 2 \cdot \diam(S \cup \set{c^*})$.
  We will show below that every candidate $c \in \hat{C}$ who is ranked ahead of $c^*$ by \emph{some} voter $v' \in S$ must be ranked ahead of $r^*$ by $v$.
  Because $v$ also ranks $c^*$ ahead of $r^*$, this implies that $\pi_{v'}(c^*) < \pi_v(r^*)$ for all $v' \in S$.
  But then, at the end of the iteration with tolerance level $\tau-1$, we would have had $S \subseteq N_{c^*}$, and therefore $|N_{c^*}| \geq p$.
  As a result, \cref{alg:URCS} would have chosen $c^*$ for some tolerance level strictly smaller than $\tau$, which would have covered $v$.
  This is a contradiction to the choice of $r^*$ as the first representative chosen in the algorithm which covers a voter in $S$.

  To prove the missing claim, let $c \in \hat{C}$ be a candidate, and $v' \in S$ a voter ranking $c$ ahead of $c^*$.
  Using the triangle inequality repeatedly, along with the assumption that $d(v, r^*) > 2 \cdot \diam(S \cup \set{c^*})$, we can bound
  
   \begin{align*}
      d(v, c) &\leq d(v, v') + d(v',c)\\
      &\leq d(v, v') + d(v', c^*)\\
      &\leq 2\cdot \diam(S \cup \set{c^*})\\
      & < d(v,r^*),
   \end{align*}
   which proves that $v$ ranks $c$ ahead of $r^*$.
\end{proof}
    Define $T := \Set{x \in V \cup C}{\rho \cdot d(x,c^*)} \leq d(x,r^*)$.
    Because the assumption \eqref{eqn:other-candidate} implies that $\rho \cdot d(v,c^*) \leq \min_{r \in R[S]} d(v, r)$, and $r^* \in R[S]$, we obtain that $S \cup \set{c^*} \subseteq T$, so $\diam(S \cup \set{c^*}) \leq \diam(T)$.
    Furthermore, because $T = A(c^*, r^*, \rho)$, \cref{lem:apollonius-diam} implies that $\diam(S \cup \set{c^*}) \leq \diam(T) \leq \frac{2}{\rho-1}\cdot d(r^*,c^*)$.
    By triangle inequality, $d(r^*,c^*) \leq d(v, r^*) + d(v, c^*) \leq (1/\rho + 1) \cdot d(v,r^*)$, again by the assumption \eqref{eqn:other-candidate}.

      Next, by substituting Claim~\ref{cl:distance-diameter}, we obtain that
      $$ \diam(S \cup \set{c^*}) \leq \frac{4}{\rho-1}\cdot\frac{\rho+1}{\rho} \cdot \diam(S \cup \set{c^*}) = 4 \cdot \frac{\rho +1}{\rho^2 -\rho} \cdot \diam(S \cup \set{c^*}). $$
    
  Solving\footnote{The implication only holds when $\diam(S \cup \set{c^*}) > 0$. However, notice that when $\diam(S \cup \set{c^*}) = 0$, \cref{cl:distance-diameter} directly implies that $d(v,r^*) = 0 \leq d(v,c^*)$, and thus $\rho \leq 1$.}
  this inequality for $\rho$ implies that $\rho \leq \frac{5+\sqrt{41}}{2} \approx 5.71$, completing the proof of the lemma.
\end{proof}

\subsection{Proof of Theorem~\ref{thm:main} and Corollaries}

\begin{proof}[Proof of Theorem~\ref{thm:main}]

Using \cref{lem:stability}, we are now ready to complete the proof of \cref{thm:main}.
In fact, we will show that the committee $R$ returned by \cref{alg:URCS} satisfies an even stronger stability guarantee than the claimed $(\alpha,\gamma)$-proportional representation.
We will show that for every coalition $S \subseteq V$ of size at least $t \cdot \alpha \cdot p$, the committee satisfies: 
\begin{align}
    \min_{\substack{R' \subseteq R\\|R'|=t}} \dsum(S, R')
& \leq \gamma \cdot \min_{\substack{C' \subseteq C\\|C'|=t}} \dsum(S, C').
  \label{eqn:stronger-representativeness}
\end{align}

That is, for the coalition $S$, there is a subcommittee $R'$ of size $t$ which is almost as good as the best $C$.
Note that the cost of $S$ for $R'$ obviously is an upper bound on the sum of costs for voters $v \in S$ for their individually optimal size-$t$ subcommittees.

Let $S$ be an arbitrary coalition of size $|S| \geq t \cdot \alpha \cdot p$.
Let $C^*$ be a set of $t$ candidates with smallest total distance to $S$, i.e., $C^* \in \argmin_{C': |C'| = t} \dsum(S, C')$. We will define a perfect matching between $C^*$ and $R$ such that for every $c \in C^*$, its match $r \in R$ is an ``approximately good'' alternative. Overall, this will demonstrate that $R$ is not much worse than $C^*$.

Let $c_1, c_2, \ldots, c_t$ be an enumeration of the candidates in $C^*$, such that the candidates in $C^* \cap R$ precede the ones in $C^* \setminus R$; apart from this requirement, the order can be arbitrary.
We define the matching representatives $r_1, \ldots, r_t$ iteratively. First, for each $c_i \in C^* \cap R$, we define $r_i = c_i$. Subsequently, for each iteration $i$,
having already defined $r_1, \ldots, r_{i-1}$, we let $S_i = S \setminus \bigcup_{j=1}^{i-1} N_{r_j}$, i.e., $S_i$ is the set of all voters in $S$ except those covered by the first $i-1$ selected candidates $r_j$.
Let $T_i \in \argmin_{T \subseteq S_i, |T| = p} \sum_{v \in T} d(v,c_i)$ be a subset of $S_i$ of size $|T_i| = p$ comprising the $p$ voters in $S_i$ closest to $c_i$ (ties broken arbitrarily).
As $|S| \geq \ceil{\alpha \cdot t \cdot p}$ and each $N_{r_j}$ has size at most $p$, $|S_i|$ has size at least $p$ for each $i$, and thus $T_i$ is well defined.
Because $c_i \notin R$ in the current case, by \cref{lem:stability}, there exists a voter $v_i \in T_i$ and candidate $r_i \in R[T_i] \subseteq R[S]$ such that
\begin{equation}
  d(v_i, r_i) \leq \rho \cdot d(v_i, c_i). \label{eq:coreset}
\end{equation}

Now, we confirm that the resulting assignment is indeed a matching. Observe that by definition of $R[S_i]$, we know that $S_i \cap N_{r_j} = \emptyset$ for all $j < i$; in particular, $r_j \notin R[S_i]$ for all $j < i$. This implies that $r_1, \ldots, r_t$ are distinct from each other, and so the assignment is a matching.
We also remark here that \cref{eq:coreset} holds for the $i$ with $c_i \in R$ as well, because $r_i = c_i$ implies $d(v_i, r_i) = d(v_i, c_i)$ for those candidates.

In the rest of the proof, we will show that $r_i$ is an approximately good alternative to $c_i$ for all of $S$, showing that $\dsum(S, r_i) < \gamma(\alpha) \cdot \dsum(S, c_i)$. Fix an arbitrary index $i \in \set{1, \ldots, t}$. Let $\hat{v}_i$ be a voter in $T_i$ maximizing the distance $d(v, c_i)$ to $c_i$ over all $v \in T_i$, i.e., a voter at (or possibly tied for) the \Kth{p} largest distance from $c_i$ among voters in $T_i$.

For any voter $v \in S$, using the triangle inequality and \eqref{eq:coreset}, we can bound the distance
\begin{align}
  d(v, r_i) 
    &\leq d(v, c_i) + d(v_i, c_i) + d(v_i, r_i) \nonumber
    \\ &\leq d(v, c_i) + (1+\rho) \cdot d(v_i, c_i) \label{eqn:for-alpha-one}
    \\ &\leq d(v, c_i) + (1+\rho) \cdot d(\hat{v}_i, c_i) \nonumber
    \\ &\leq d(v,c_i) + (1+\rho) \cdot \max\set{d(v, c_i), d(\hat{v}_i, c_i)}.\nonumber
\end{align}
  Let $P_i = \Set{v \in S}{d(v,c_i) < d(\hat{v}_i, c_i)}$ be the subset of $S$ containing all voters $v$ with $d(v, c_i) < d(\hat{v}_i, c_i)$.
  Because $T_i$ was chosen to be the $p$ voters closest to $c_i$ inside $S_i$, and $\hat{v}_i$ the voter furthest from $c_i$ in $T_i$, we get that $P_i \subseteq T_i \cup \bigcup_{j=1}^{i-1} N_{r_j}$ and so $|P_i| \leq p \cdot i$.
Summing the distances to $r_i$ over all voters $v \in S$, and using (in the second step) that $d(\hat{v}_i, c_i)\leq d(v, c_i)$ for all $v \in S \setminus P_i$, we obtain the bound
\begin{align*}
  \sum_{v \in S} d(v, r_i) 
   &\leq {\sum_{v \in S} d(v,c_i) + (1+\rho)} \cdot \left(\sum_{v \in P_i} d(\hat{v}_i, c_i) + \sum_{v \in S \setminus P_i} d(v, c_i)\right)\\
   &\leq {\sum_{v \in S} d(v,c_i) + (1+\rho)} \cdot \left( \left(\frac{|P_i|}{|S| - |P_i|} + 1\right) \cdot \sum_{v \in S \setminus P_i} d(v, c_i)\right)\\
   &\leq {\sum_{v \in S} d(v,c_i) + (1+\rho)} \cdot \left( \left(\frac{p \cdot i}{|S| - p \cdot i} + 1\right) \cdot \sum_{v \in S \setminus P_i} d(v, c_i)\right)\\
   &\leq {\sum_{v \in S} d(v,c_i) + (1+\rho)} \cdot \left( \left(\frac{1}{\alpha - 1} + 1\right) \cdot \sum_{v \in S \setminus P_i} d(v, c_i)\right)\\
   &\leq {\left(1 + (1+\rho) \cdot \frac{\alpha}{\alpha-1} \right)} \cdot \sum_{v \in S} d(v, c_i)\\
\end{align*} 
In the penultimate step, we used that $\frac{p \cdot i}{|S| - p \cdot i} \leq {\frac{p \cdot i}{\alpha \cdot p \cdot i - p \cdot i}} = \frac{1}{\alpha - 1}$. Finally, summing up over all indices $i$, we obtain
\[
  \sum_{i=1}^t \dsum(S, r_i)
  \leq {\left( 1 + (1 + \rho) \cdot \frac{\alpha}{\alpha-1} \right)} \cdot \sum_{i=1}^t \dsum(S, c_i).
\]
Substituting $\rho = \frac{5+\sqrt{41}}{2}$ now completes the proof.
\end{proof}

As a special case of \cref{thm:main} and \cref{thm:no-augmentation}, we obtain \cref{cor:core-fairness}, simply by focusing on just the case $t=1$ in \cref{thm:main}.
In fact, by recalling that the proof of \cref{thm:main} established the somewhat stronger guarantee \cref{eqn:stronger-representativeness}, we obtain the following \cref{cor:single}, strengthening \cref{cor:core-fairness}.
This corollary states that no coalition of at least $\alpha \cdot p$ voters (for $\alpha > 1$) has a candidate outside $R$ whom they strongly prefer \emph{on average} to their \emph{best single} candidate in $R[S]$:

\begin{corollary}
    \label{cor:single}
    Let $R$ be the committee output by the \ear.
    Let $\alpha > 1$, and $\beta(\alpha) = {1 + \frac{7+\sqrt{41}}{2}\cdot \left(\frac{\alpha}{\alpha-1}\right) \approx 1 + 6.71 \cdot \left(\frac{\alpha}{\alpha-1}\right)}$.
    For any coalition $S$ of size $|S| \geq \alpha \cdot p$, 
    \begin{align*}
      \min_{r \in R[S]} \dsum(S, r)
      & \leq \beta(\alpha) \cdot \min_{c \in C\setminus R} \dsum(S, c).
    \end{align*}
\end{corollary}

Finally, we show how \cref{thm:main} implies \cref{cor:single-winner-distortion}, i.e., the existence of an extremely simple single-winner voting rule with constant distortion.

\begin{proof}[Proof of \cref{cor:single-winner-distortion}]
Consider the committee $R$ output by \cref{alg:URCS} when run with $k=2$.
By \cref{thm:main}, applied with $\alpha = n/p \leq 2$, we get that $ \min_{r \in R} \dsum(V, r) \leq \gamma(2) \cdot \min_{c \in C} \dsum(V, c)$.
This implies that at least one representative in $R$ has distortion at most $\gamma(2) \approx 14.42$ for the single-winner election.
Let $r_1$ be the winner of the majority election between the two candidates in $R$, and $r_2$ the other candidate. Since $r_1$ is preferred over $r_2$ by at least half of the voters, Lemma~6 of \citet{anshelevich:bhardwaj:elkind:postl:skowron} implies that 
$\dsum(V, r_1) \leq 3 \dsum(V, r_2)$.
This implies that $r_1$ has distortion at most $3 \cdot \gamma(2) \leq 44$ for the single-winner election. 
\end{proof}

\subsection{Proportional Representation Without Augmentation: Proof of \cref{thm:no-augmentation}}
\label{sec:without_augmentation}

Let $S$ be a coalition of size $|S| = p \cdot t$.
Consider the $r_i, c_i, v_i, i = 1, \ldots, k$ defined in the proof of \cref{thm:main}.
Using the bound from \cref{eqn:for-alpha-one} that $d(v, r_i) \leq d(v, c_i) + (1+\rho) \cdot d(v_i, c_i)$ for all $v \in S$ and all $i$, we obtain that
\begin{align}
\dsum(S, R)
  & \leq \sum_{v \in S} \sum_i (d(v, c_i) + (1+\rho) \cdot d(v_i, c_i))
    = \dsum(S,C^*) + (1+\rho) \cdot |S| \cdot \sum_i d(v_i, c_i).
    \label{eqn:algo-cost-non-augmented}
\end{align}

The rest of the proof will be concerned with upper-bounding $\sum_i d(v_i, c_i)$.
Notice that it is possible that some/many of the $v_i$ are the same, necessitating the following approach.
For each $i < k$, let $u_i \in S_i \setminus T_i$ such that all of the $u_i$ are distinct.
Notice that because $|S_i| \geq (k+1-i)$ for all $i$, it is possible to define such $u_i$ inductively by decreasing $i$, starting with $i=k-1$: there is always some $u \in S_i$ available which was not chosen as $u_j$ for any $j > i$.
Further define $u_k = v_k$.
First, observe that each $v$ appears at most twice\footnote{The only such $v$ is possibly $v=u_k = u_{k-1}$, and that only in the case $p=1$. Otherwise, each $v$ appears at most once.} as a $u_i$.
Second, because of the choice of $T_i$ as being closest to $c_i$ among voters in $S_i$, and because $v_i \in T_i$, we get that $d(u_i, c_i) \geq d(v_i, c_i)$ for all $i$, so $\sum_i d(v_i, c_i) \leq \sum_i d(u_i, c_i)$.
Because each $u_i$ appears at most twice, we immediately get the bound that $\sum_i d(u_i, c_i) \leq 2\dsum(S, C^*)$. However, to prove our desired guarantee, we need the stronger bound that $\sum_i d(u_i, c_i) \leq O(1/k) \cdot \dsum(S, C^*)$. The remainder of the proof will be concerned with proving this bound.
  
We start from the inequality $2 \sum_{v \in S} \sum_i d(v,c_i) \geq \sum_{i,j} d(u_i,c_j)$, which also follows because each $v$ appears at most twice as a $u_i$.
We will now lower-bound the sum on the right-hand side.
We begin by defining the following sets of pairs of indices $(i, j)$ with $i, j = 1, \ldots, k$:
\begin{align*}
   A & = \Set{(i,j)}{d(u_i, c_j) \geq \frac{d(u_i, c_i)}{2} \text{ and } d(u_j, c_i) \geq \frac{d(u_j, c_j)}{2}}
\\ B & = \Set{(i,j)}{d(u_i, c_j) \geq \frac{d(u_i, c_i)}{2} \text{ and } d(u_j, c_i) < \frac{d(u_j, c_j)}{2}}
\\ C & = \Set{(i,j)}{d(u_i, c_j) < \frac{d(u_i, c_i)}{2} \text{ and } d(u_j, c_i) \geq \frac{d(u_j, c_j)}{2}}
\\ D & = \Set{(i,j)}{d(u_i, c_j) < \frac{d(u_i, c_i)}{2} \text{ and } d(u_j, c_i) < \frac{d(u_j, c_j)}{2}}.
\end{align*}

When $d(u_i, c_j) < \frac{d(u_i, c_i)}{2}$, by triangle inequality, we have that
$d(c_i, c_j) \geq d(u_i, c_i) - d(u_i, c_j) > \frac{d(u_i, c_i)}{2}$.
Again by triangle inequality, this implies that for all $v$, we have that $d(v, c_i) + d(v, c_j) \geq d(c_i, c_j) > \frac{d(u_i, c_i)}{2}$.
Similarly, when $d(u_j, c_i) < \frac{d(u_j, c_j)}{2}$, we have that $d(v,c_i) + d(v,c_j) > \frac{d(u_j,c_j)}{2}$ for all $v$.
Now consider the four classes of pairs of indices defined above:

\begin{enumerate}
\item If $(i,j) \in A$, then $d(u_i, c_j) + d(u_j, c_i) \geq \frac{1}{2} \cdot (d(u_i, c_i) + d(u_j, c_j))$, directly from the definition of $A$.
\item If $(i,j) \in B$, then $d(v,c_i) + d(v,c_j) > \frac{d(u_j,c_j)}{2}$ for all $v$. In particular for $v=u_i$, we get that $d(u_j,c_j) \leq 2d(u_i, c_i) + 2d(u_i, c_j) \leq 6d(u_i,c_j)$ because $d(u_i,c_j) \geq \frac{d(u_i,c_i)}{2}$. Because we have $d(u_i, c_j) \geq \frac{d(u_i,c_i)}{2}$ and $d(u_i, c_j) \geq \frac{d(u_j,c_j)}{6}$, by adding the first inequality plus three times the second, we get that $d(u_i, c_j) + d(u_j, c_i) \geq d(u_i, c_j) \geq \frac{1}{8} \cdot (d(u_i, c_i) + d(u_j, c_j))$.
\item If $(i,j) \in C$, then a completely symmetric argument with the roles of $i,j$ reversed gives that $d(u_i, c_j) + d(u_j, c_i) \geq \frac{1}{8} \cdot (d(u_i, c_i) + d(u_j, c_j))$.
\item Finally, when $(i,j) \in D$, we have that $d(v,c_i) + d(v,c_j) \geq \frac{d(u_i, c_i)}{2}$ and $d(v,c_i) + d(v,c_j) \geq \frac{d(u_j, c_j)}{2}$ for all $v$. By averaging these two inequalities, we get that
\begin{align}
  d(v,c_i) + d(v,c_j) & \geq \frac{1}{4} \cdot (d(u_i, c_i) + d(u_j, c_j)). \label{eqn:case-D}
\end{align}
\end{enumerate}

Notice that in the three first cases, we have that $d(u_i, c_j) + d(u_j, c_i) \geq \frac{1}{8} \cdot (d(u_i, c_i) + d(u_j, c_j))$, so summing up over all pairs $(i,j)$, we obtain that
\begin{align}
  2 \sum_{i,j} d(u_i, c_j)
  & \geq \sum_{(i,j) \notin D} d(u_i, c_j) + d(u_j, c_i)
  \geq \frac{1}{8} \cdot \sum_{(i,j) \notin D} d(u_i,c_i) + d(u_j, c_j). \label{eqn:case-non-D-sum}
\end{align}

We now focus on the fourth case.
Note that $(i,j) \in D$ if and only if $(j,i) \in D$.
Consider an undirected graph $G$ on $\set{1, \ldots, k}$ which contains the edge $(i,j)$ if and only if $(i,j) \in D$.
Because the maximum degree of any node in $G$ is at most $k-1$, by Vizing's Theorem\footnote{Vizing's Theorem states that any simple undirected graph can be edge-colored with at most one more color than its highest degree, ensuring no two adjacent edges share the same color.}, the edges of $G$ can be partitioned into $k$ matchings (some of which may be empty). 
Let $M_{\ell}$ for $\ell = 1, \ldots, k$ be the pairs $(i,j) \in D$ in the \Kth{\ell} matching.
Substituting $v=u_{\ell}$ in \cref{eqn:case-D}, we get for any pair\footnote{This holds for all $(i,j) \in D$, not just those in $M_{\ell}$. But we will use it only for $(i,j) \in M_{\ell}$.} $(i,j) \in M_{\ell}$ that
$d(u_{\ell}, c_i) + d(u_{\ell}, c_j) \geq \frac{1}{4} \cdot (d(u_i,c_i) + d(u_j, c_j))$.
Summing over all $\ell = 1, \ldots, k$ and all $(i,j) \in M_{\ell}$, we obtain that
\[ \sum_{\ell=1}^k \sum_{(i,j) \in M_{\ell}} d(u_{\ell},c_i) + d(u_{\ell}, c_j)
    \geq \frac{1}{4} \cdot \sum_{(i,j) \in D} d(u_i,c_i) + d(u_j,c_j).
\]
  
On the left-hand side, observe that because each $M_{\ell}$ is a matching, and $u_{\ell} \neq u_{\ell'}$ for all $\ell \neq \ell'$, each term $d(u_{\ell}, c_i)$ can occur at most once.
In other words, we get that
$\sum_{\ell=1}^k \sum_{(i,j) \in M_{\ell}} d(u_{\ell},c_i) + d(u_{\ell}, c_j) \leq \sum_{i,j} d(u_i, c_j)$, where the right-hand sum is \emph{not} restricted to $D$.
  In summary, we have shown that
\begin{align}
  \sum_{i,j} d(u_i, c_j) & \geq \frac{1}{4} \cdot \sum_{(i,j) \in D} d(u_i, c_i) + d(u_j, c_j). \label{eqn:case-D-sum}
\end{align}

Adding twice \cref{eqn:case-non-D-sum} and \cref{eqn:case-D-sum}, we get that
\[
 5 \cdot \sum_{i,j} d(u_i, c_j)
  \geq \frac{1}{4} \cdot \sum_{(i,j)} d(u_i, c_i) + d(u_j, c_j)
   = \frac{k}{2} \cdot \sum_i d(u_i, c_i)
  \geq \frac{k}{2} \cdot \sum_i d(v_i, c_i),
\]
so $\sum_i d(v_i, c_i) \leq \frac{10}{k} \cdot \sum_{i,j} d(u_i, c_j) \leq \frac{20}{k} \cdot \dsum(S, C^*)$.
Substituting this bound into \cref{eqn:algo-cost-non-augmented}, we finally obtain that
\[
  \dsum(S, R) \leq \dsum(S, C^*) + (1+\rho) \cdot |S| \cdot \frac{20}{k} \cdot \dsum(S, C^*)
  = O(n/k) \cdot \dsum(S, C^*).
\]

This completes the proof that the output of \cref{alg:URCS} is $(1,O(n/k))$-proportionally representative. \qed

\section{Proportional Representation with Known Distances}
\label{sec:known-distances}

When the metric space is known, i.e., in the cardinal information model, we provide an algorithm with better approximation constants.
In particular, we establish that a slight modification of the \textsc{Greedy Capture} algorithm of \citet{pfc} provides good $(\alpha, \gamma)$-representativeness guarantees.

\gc works as follows: it continuously grows balls around each candidate, at the same rate.
When a ball around a candidate $c$ contains $p = \ceil{n/k}$ voters who have not been removed yet, $c$ is included in the committee, and the voters are removed.
Even once $c$ is included in the committee, the ball around $c$ continues growing, and any additional voter included in $c$'s ball is immediately removed as well.
In the nomenclature of our \cref{alg:URCS}, this means that candidate $c$ can ``cover'' more than $p$ voters under \gc.

We modify the algorithm by preventing balls of included candidates from growing.
That is, as soon as $c$ is included in the committee, and the voters in the ball are removed, $c$ and its ball are removed from further consideration.
The full algorithm is given as \cref{alg:CARDURC}.

\begin{algorithm}[H] 
    \caption{\textsc{Truncated Greedy Capture} \label{alg:CARDURC}}
    \textbf{Input: } $(V,C,k)$, and metric $d$ on $V \cup C$\\
    \textbf{Output: } Committee $R$
    \begin{algorithmic}
        \State Let $U \leftarrow V$ be the initial set of uncovered voters.
        \State Let $R \leftarrow \emptyset$ be the selected committee.
        \State Let $\delta = 0$ be the ball radius.
        \State Set $N_c \leftarrow \emptyset$ for each candidate $c \in C$.

        \While{$U \neq \emptyset$}
            \State Continuously increase $\delta$. 
            \For{$v \in V$ in arbitrary order}
              \If{$v \in U$}
                \For{$c \in C \setminus R$ in arbitrary order}
                    \If{$d(v,c) \leq \delta$}
                      \State Set $N_c \leftarrow N_c \cup \set{v}$.
                        \If{$|N_c| = \ceil{n/k}$}
                           \State $R \leftarrow R \cup \set{c}$, i.e., include $c$ in the committee.
                           \State $N_{c'} \leftarrow N_{c'} \setminus N_c$ for all $c' \in C \setminus R$, i.e., update the neighborhood for non-committee candidates.
                           \State $U \leftarrow U \setminus N_c$, i.e., update the set of uncovered voters.
                           \State We say $N_c$ has been covered by $c$.
                        \EndIf
                     \EndIf
                  \EndFor
               \EndIf
            \EndFor
        \EndWhile
        \State \textbf{if} $|R|<k$ \textbf{then} add $k-|R|$ arbitrary candidates to $R$.
    \end{algorithmic}
\end{algorithm}

\begin{theorem} \label{thm:distances-known}
  When all distances are known to the algorithm, \textsc{Truncated Greedy Capture} yields a committee $R$ which is simultaneously $(\alpha, \gamma(\alpha))$-proportionally representative for all $\alpha > 1$, with 
    $\gamma(\alpha) = {1 + (2+\sqrt{2})\cdot \frac{\alpha}{\alpha-1} \approx 1 + 3.42 \cdot \frac{\alpha}{\alpha-1}}$.
\end{theorem}

In order to prove \cref{thm:distances-known}, we first prove that \cref{alg:CARDURC} satisfies the analogue of \cref{lem:stability}, but with improved constants.
Recall that for any coalition $S$ of voters, $R[S] \subseteq R$ denotes the set of candidates $r \in R$ who had at least one voter from $S$ in their neighborhood when $r$ was added to $R$.

\begin{lemma}
    \label{lem:stability-cardinal}
    The committee $R$ output by \cref{alg:CARDURC} satisfies the following stability property, with $\rho=1+\sqrt{2} \approx 2.42$.
    For any coalition $S$ of size $|S| \geq p = \ceil{n/k}$, there exists a voter $v \in S$ with 
    $$ \min_{r \in R[S]} d(v,r) \leq \rho \cdot \min_{c \in C \setminus R} d(v,c).$$
\end{lemma}

The proof of this lemma is very similar to the proof of Theorem~1 of \citet{pfc}.
The fact that balls around included candidates $c$ continue to grow in the original \gc algorithm turns out to be unimportant for its theoretical guarantees (though likely beneficial in practice).
By stopping the growth, we obtain the stronger guarantee of the existence of an $r \in R[S]$ (rather than $r \in R$).
For completeness, we give a self-contained proof of \cref{lem:stability-cardinal}.

\begin{proof}[Proof of \cref{lem:stability-cardinal}]
  Let $R$ be the output of \cref{alg:CARDURC}.
  Let $S \subseteq V$ be any coalition of voters of size $|S| \geq p$.
  Let $c \in C \setminus R$ be an arbitrary candidate.
  We will show the existence of a candidate $r \in R[S]$ with $d(v,r) \leq \rho \cdot d(v,c)$.
  
  Let $\delta = \max_{v \in S} d(v, c)$, which implies that $S \subseteq B(c, \delta)$.
  There must exist some candidate $r \in R[S]$ such that $B(r, \delta) \cap S \neq \emptyset$;
  otherwise, when \cref{alg:CARDURC} considered ball radius $\delta$, it would have included $c$ (as all voters in $S$ would have been uncovered and in $N_c$ at that time).
  Notice that the slight strengthening of our result compared to Theorem~1 of \citet{pfc} is that they only guarantee $r \in R$, whereas we guarantee (and need) $r \in R[S]$.

  Fix a candidate $r$ with $B(r, \delta) \cap S \neq \emptyset$ (whose existence we just proved).
  Let $v \in B(r, \delta)$ be arbitrary, and $v^*$ be a voter with $d(v^*, c) = \delta$.
  Using the triangle inequality, along with the bounds that $d(v^*,c) = \delta$ and $d(v,r) \leq \delta$, we now show that at least one of $v, v^*$ prefers $c$ over $r$ by at most a small amount:
  \begin{align*}
    \min \left(\frac{d(v,r)}{d(v,c)}, \frac{d(v^*,r)}{d(v^*,c)}\right)
    & \leq \min \left(\frac{d(v,r)}{d(v,c)},
                    \frac{d(v^*,c) + d(v,c) + d(v,r)}{d(v^*,c)}\right)
    \\ & \leq \min \left(\frac{\delta}{d(v,c)}, 2 + \frac{d(v,c)}{\delta} \right)
    \\ & \leq \max_{z \geq 0} \min\set{z,2+\frac{1}{z}}
    \\ & = 1 + \sqrt{2}.
  \end{align*}
  With the voter for the given set $S$ being either $v$ or $v^*$, this completes the proof.
\end{proof}

\begin{proof}[Proof of \cref{thm:distances-known}]
In the proof of \cref{thm:main}, the only properties of \cref{alg:URCS} that were used were:

\begin{itemize}
\item \cref{lem:stability}.
\item The disjointness of the neighborhoods $N_r$ for $r \in R$.
\item The fact that each neighborhood $N_r$ for $r \in R$ had size $|N_r| = p$.
\end{itemize}

Because \cref{alg:CARDURC} also satisfies the second and third properties, we can
substitute the improved bound of \cref{lem:stability-cardinal} into the proof and adjust the constants; this completes the proof of \cref{thm:distances-known}.
\end{proof}

Finally, we briefly analyze the running time of \cref{alg:CARDURC}.
While the algorithm is stated as running in continuous time, the points $\delta$ at which a ball includes another voter can be easily pre-computed (they comprise exactly the distances $d(v,c)$ for all pairs $(v,c)$) and sorted in increasing order.
All checks can then be performed by simple operations or set membership tests, and the remaining analysis is exactly as for \cref{alg:URCS}, giving a running time of $\tilde{O}(nm)$, i.e., essentially linear unless $n \gg m$.

\section{Related Work} \label{sec:related-work}

\subsection{Multi-Winner Metric Distortion}

Several papers have proposed extensions of the notion of (metric) distortion to multi-winner (or committee) elections.
\citet{goel-relating-metric-and-fairness} consider the cost of a set $R$ to be the sum of all distances between $V$ and the members of $R$.
Because this sum decomposes, it is minimized by choosing the $k$ candidates individually closest to $V$ (in terms of sums of distances); as a result, the set $R$ tends to be as ``homogeneous'' as possible.
For example, if $V$ is partitioned into two clusters of 51\% and 49\%, and there are enough candidates near each cluster, then the optimal solution would choose \emph{all} candidates close to the cluster comprising 51\% of $V$.
We note that \citet{goel-relating-metric-and-fairness} did not propose this objective as a natural objective for evaluating how representative $R$ is; rather, the goal was to show that many results relating (and bounding) distortion and fairness naturally generalize from single-winner to multi-winner elections.

An alternative notion was proposed by \citet{metric-multi}. Their definition is parametrized by $q \leq k$: each individual $v \in V$ evaluates the cost of $R$ as the cost of the \Kth{q} closest representative in $R$; the objective to minimize is then the sum of these costs.
When $q=1$, the objective coincides with (uncapacitated) $k$-median; in contrast, for $q=k$, a committee has low cost for $v$ only if all of its members are close to $v$. Thus, in the regime of large $q$, the definition suffers from the same drawback as that of \citet{goel-relating-metric-and-fairness}: it rewards committees $R$ (almost) all of whose members are close to the largest cluster within $V$.
\citet{metric-multi} show that while the objective can be well approximated with ordinal information in the ``homogeneous'' case $q > k/2$, the distortion is unbounded for $q < k/3$; though no results are shown under resource augmentation.

This notion of $q$-cost was used as a definition of fairness or representativeness in the context of sortition in subsequent work by \citet{sortition-representative-fair} and \citet{ebadian-micha-sortition}.
In \citet{sortition-representative-fair} and part of \citet{ebadian-micha-sortition}, this notion is used primarily as an analysis tool: the analysis focused on uniformly random selection of individuals, and the representativeness/fairness of the resulting panels, i.e., no knowledge of the metric whatsoever (not even ordinal) was assumed. This is because the random selection procedure itself was a primary focus of the work.
However, in addition, \citet{ebadian-micha-sortition} were also interested in the question of how much more representativeness of panels could be achieved with full knowledge of the underlying metric; we discuss those results and their relationship with our work in \cref{sec:related-clustering}.

\subsection{The Core and Proportionality in Social Choice}
\label{sec:related-core-proportionality}
There has been a large amount of prior work relating notions of fairness, multi-winner elections, distortion, and the core. 
For example, \citet{ebadian:kahng:peters:shah:optimized-distortion} study the \emph{utilitarian distortion} of \emph{randomized} single-winner voting rules.
They use fairness both as a tool to derive novel low-distortion randomized voting rules, and --- under a different notion of proportional fairness --- as an optimization goal in its own right.
They show that $\alpha$-approximate proportional fairness in their definition implies membership in the $\alpha$-core.

In general cooperative games, the notion of core captures a stability desideratum: that the outcome be stable against deviations by any subgroup of players seeking better utilities.
The exact coalitions and available ``outside options'' of utilities give rise to specific core solution concepts. This concept is further extended to inverse utility games, where the objective is to reduce the cost incurred by the players \citep{tamir-core-location, pfc}. The stability found in the core of cost-sharing games closely mirrors the notion of proportional representation of points in a metric space. In a vein similar to proportional representation, the core in cost-sharing games guarantees that large fractions of players have low cost under the chosen solution, deterring them from deviating to alternatives.

In social choice, core definitions often emphasize proportionality, suggesting that a $\theta$ fraction of the population should ``control'' an equal fraction of the outcome \citep{moulin:fair-division}. 
In the context of electing a committee, various notions of core stability require that every sufficiently cohesive and large voter group should sufficiently ``approve'' of the committee.
Early research by \citet{dummett-voting-procedures} aimed to ensure proportionality for solid coalitions (PSC) on ranked ballots. These coalitions consist of voter groups whose set of top candidates are the same (for corresponding set sizes), though their ordering of these top candidates may differ. 

Subsequently, more research focused on approval ballots instead of ranked ballots; here, each voter specifies for each candidate whether they approve of the candidate or not; equivalently, the voter specifies the set of all approved candidates.
\citet{aziz-justified-representation} introduced the concept of Justified Representation (JR).
Justified Representation (JR) requires that cohesive groups be well-represented in the committee as follows: if a set of $\theta n$ voters is \emph{cohesive} in the sense of having at least $\theta k$ candidates whom they all approve, then at least $\theta k$ of those unanimously approved candidates must be selected. Unfortunately, a committee achieving JR may not exist.
Subsequent work of \citet{fernandez-proportional-justified-representation, full-justified-representation, skowron:proportionality-degree} further refined this concept, in part with an eye towards ensuring existence of a committee satisfying the definition. They consider various alternate and subtly differing definitions of cohesiveness; see also the survey by \citet{multiwinner-approval}.

In studying axiomatic approaches for proportional representation, it has been observed repeatedly (see the in-depth discussion of this strand of literature in \citet{brill-peters-robust-and-verifiable-proportionality}) that various cohesiveness definitions for coalitions --- both for approval and ranked ballots --- are very demanding. As a result, only few coalitions are cohesive, imposing few requirements on a selected committee.
In response, \citet{brill-peters-robust-and-verifiable-proportionality} propose new definitions for both approval and ranked ballots.
Their PJR+ notion requires the following. For any coalition $S$ which, based on proportionality, is ``entitled'' to $\ell$ committee members, if the union of committee members approved by at least one voter in the coalition has size smaller than $\ell$, then the committee is not allowed to omit any candidate who is \emph{unanimously} approved by the coalition. The rank-PJR+ notion for ranked ballots requires this to hold for the ``approval ballots'' obtained by truncating all ranked ballots at position $r$, for all $r$.
This constitutes a more demanding notion of proportional representation under ranked choice ballots, compared to PSC \citep{dummett-voting-procedures}.
\citet{brill-peters-robust-and-verifiable-proportionality} show that while Single-Transferable Vote fails rank-PJR+, the \ear satisfies it.

General notions of stability are also studied by \citet{cheng-jiang-munagala-wang:group-fairness,jiang-approximately-stable-committee}.
\citet{cheng-jiang-munagala-wang:group-fairness} again consider deviation threats in which coalitions $S$ may deviate to a set of size $\frac{|S|}{n} \cdot k$ if there exists such a set that they prefer over the chosen committee.
They consider both approval and ranked ballots: for approval ballots, voters prefer the committee that contains more candidates whom the voter approves, while for ranked ballots, voters prefer the committee with the single highest-ranked candidate according to the voter's order. The main focus of \citet{cheng-jiang-munagala-wang:group-fairness} is on lotteries over committees that achieve stability in this sense (since no deterministic stable committee may exist).
\citet{jiang-approximately-stable-committee} consider more general monotone preferences of voters over committees. They consider approximate stability, similar to our resource augmentation: they require that no coalition $S$ have a deviation threat of size $\frac{|S|}{cn} \cdot k$ for $c \geq 1$ that they prefer over the chosen committee. The main contribution of \citet{jiang-approximately-stable-committee} is to show that for a constant $c$, there always exists an approximately stable committee in this sense, so long as the voters' preferences are monotone. They prove this by starting from a lottery over committees, which is then derandomized.

\subsection{Clustering, Fairness and Proportionality}
\label{sec:related-clustering}
Representing a large point set is also a --- sometimes explicit, sometimes implicit --- goal of clustering points in a metric space.
A review of this very large literature is beyond the scope of our work.
Of the common objective functions (most notably, $k$-center, $k$-means, and $k$-median), $k$-median \citep{AGKMMP} is closest to our objective here, since it minimizes the sum of distances of points to the respective closest of the selected $k$ cluster centers.
In the basic definition, like proportionally fair clustering, it suffers from the fact that a large and dense cluster can be ``served'' by just one selected representative.

Capacitated versions address this issue by limiting the number of points ``served'' by a cluster center (see, e.g., \citet{byrka:fleszar:rybicki:spoerhase}). 
Our algorithms for both ordinal and cardinal models similarly assign to each representative $c$ a subset of size $p$ from $V$ that $c$ ``covers''.
To appreciate the difference between the objectives, suppose that roughly $p$ points are at a large distance $M$ from all others points which are within a unit circle. 
If a representative is at distance $(1-\epsilon) M$ from these remote points, a $k$-median solution might include it since the objective reduction by $\epsilon M$ would outweigh any decisions made within the unit circle. 
In contrast, our proportional representation objective might exclude it since the relative improvement would be marginal. 
Thus, the proportional representation objective is much less sensitive to few outliers. 
It should also be noted that both objectives require resource augmentation for positive results \citep{byrka:fleszar:rybicki:spoerhase}. 
While the objectives look quite different, it would be interesting to explore a link between them and to determine if one implies non-trivial guarantees for the other.

Notions of fairness have also been incorporated into clustering problems more explicitly. These can be roughly divided into fairness notions towards individuals and towards coalitions.
A notion of fairness towards individuals was proposed by \citet{jung-individual-fairness,chakrabarti-equitable-k-center}: they consider a committee of size $k$ to be $\alpha$-fair if for each individual $i$, the committee contains a member that is no more than $\alpha$ times as far from $i$ compared to $i$'s \Kth{n/k} closest candidate. Fairness in this sense may come at a significant cost for the overall utility function (such as $k$-median), and \citet{mahabadi:vakilian:individual-fairness,vakilian:yalciner:individual-fairness} give algorithms with improved tradeoffs between the objectives of fairness and overall clustering quality.

An alternative notion of individual fairness was considered by \citet{chakrabarti-equitable-k-center}. Their notion somewhat resembles envy-freeness: each point has a self-declared set of ``peers'', and would like its distance to its closest center to be larger by at most a factor $\alpha$ than that of its peers; the latter could be the minimum or average distance of peers to their closest center.

Yet another notion was considered in the work of \citet{ebadian-micha-sortition}. Their work was motivated by \emph{Sortition}: choosing random citizen panels such that each citizen has (approximately) uniform probability of being part of the panel, and the panel overall is representative in the sense of multi-winner elections discussed above: i.e., for each individual, the panel contains $q$ members approximately closest to the individual. In order to obtain a good tradeoff between individual inclusion probabilities and approximation to the core, \citet{ebadian-micha-sortition} consider another modified version of the \textsc{Greedy Capture} algorithm. Like our \textsc{Truncated Greedy Capture} algorithm, their version freezes a ball when representatives are chosen; however, this happens only when a ball contains $q$ times the Hare quota many points, and at this point, the $q$ representatives from the ball are chosen uniformly at random.
Interestingly, the approximation to the core is shown to be the same $\frac{5+\sqrt{41}}{2}$ which appears in our analysis of the \ear for Proportionally Fair Clustering. The technical steps leading to this bound (i.e., specific triangle inequalities and combinations) are similar, even though the algorithms and overall analysis are different.

An alternative approach, more in line with our definition, is to require fairness (or lack of deviation incentives) for sufficiently large groups.
Perhaps the most prominent examples are the notions of proportional fairness \citep{pfc,pfc-revisited} and approximate core \citep{LLSWW:core-fairness}, both discussed in depth in \cref{sec:pr-vs-pfc}.

Simultaneously and independently of our work, \citet{aziz-lee-chu-vollen-proportional-clustering} also propose a notion of proportional representation.
Their motivation is quite similar to ours: they also observe that neither proportionally fair clustering nor the approximate core achieve the goal (stated in those works) of representing clusters proportionally to their size. 
They formulate a basic axiom called \textsc{Unanimous Proportionality (UP)}, which requires that if there is a coalition $S$ comprising a $\theta$ fraction of the population all of whose members are in the \emph{exact same} location, then a $\theta$ fraction of the committee should be made up of points closest to $S$.
This is a clean axiomatization of the issue we discuss in \cref{sec:pr-vs-pfc}; indeed, our notion of proportional representation also leads to satisfying the UP axiom.
\citet{aziz-lee-chu-vollen-proportional-clustering} then propose a definition they term \textsc{Proportionally Representative Fairness (PRF)}.
This notion requires the following, for each set $S$. Let $y$ be the diameter of $S$, and $\ell$ the number of candidates who are at distance at most $y$ from all members of $S$. Let $\theta$ the fraction of the population that $S$ comprises. Then, the committee must contain at least $\min(\theta k, \ell)$ candidates who are at distance at most $y$ from all members of $S$. 
\citet{aziz-lee-chu-vollen-proportional-clustering} then propose an algorithm for selecting a committee satisfying their definition. The algorithm is essentially an \ear with known distances; that is, balls around candidates are grown, and when a ball contains $n/k$ voters, the candidate is included, and the voters are ``removed'' (by having their weight decreased). As with the algorithms we consider, proportionality is ensured by having the balls around included candidates not grow any further.
Notice that the algorithms are designed with knowledge of the metric space.
The notions of proportional representation of our work and \citet{aziz-lee-chu-vollen-proportional-clustering} are obviously similar in their goals and approaches, although we are not aware of a formal reduction.
It should be noted that while our notion is a strict strengthening of the notion of core fairness proposed by \citep{LLSWW:core-fairness}, this is not the case for the notion of \citet{aziz-lee-chu-vollen-proportional-clustering}.

In further simultaneous and independent work, \citet{kellerhals-peters-proportional-fairness-in-clustering} aim to draw connections between the plethora of different fairness/proportionality notions alluded to above.
Among others, they show that proportional fairness and individual fairness imply each other (up to losses of constant factors).
More importantly, they define metric notions of ranked fairness. Recall from \cref{sec:related-core-proportionality} that notions of justified representation (JR, EJR, PJR, PJR+) can be extended to ranked (rather than approval) ballots by considering all the cutoffs $r$, and requiring the conditions to hold for each implied approval ballot.
As an alternative, \citet{kellerhals-peters-proportional-fairness-in-clustering} propose notions of rank-JR, rank-PJR, rank-PJR+ based on \emph{distance} cutoffs: they consider agents as approving all candidates within distance $r$, and require that the corresponding justified representation axioms hold for the implied approval ballots.
They show that various natural algorithms (such as \textsc{Greedy Capture} and a distance-based \textsc{Expanding Approvals Rule}) satisfy corresponding axioms. Furthermore, they show implications between these new axioms and proportional and individual fairness (as well as the proportional representation notion of \citet{aziz-lee-chu-vollen-proportional-clustering}), obtaining results on the representation guarantees for the outputs of these algorithms for various notions. In several cases, the obtained guarantees match the best known results.

\section{Concluding Remarks and Open Questions}
\label{sec:conclusion}

We presented a natural new definition of when a committee is ``representative'' of a larger set of points in a metric space.
This definition is applicable for documents (where the metric is typically known) or for the election of a committee or other representative political body.
In the latter case, typically, only ordinal information about the metric is available, and our definition gives a natural objective for distortion-based analysis of multi-winner elections.
Our main result is that the \ear of \citet{aziz:lee:expanding-approvals-rule} achieves constant representativeness with resource augmentation by a constant factor, even with just ordinal information;
we also showed that some resource augmentation in the analysis is unavoidable, as for large committees, there are examples for which a polynomial lower bound on representativeness is unavoidable.

\citet{aziz:lee:expanding-approvals-rule} showed that the \ear satisfies several desirable axiomatic properties for multi-winner elections, in addition to being a natural rule in its own right.
We believe that our result thus adds to the evidence for \ear being a potentially useful rule to be used in practice.

\cref{thm:main,thm:distances-known} give bounds for general metric spaces.
When the metric is known to be Euclidean, by using the stronger bound of Theorem~5 of \citet{pfc-revisited} in place of \cref{lem:apollonius-diam} (Theorem~4 of \citet{pfc-revisited}), the constant in \cref{lem:stability} can be improved, which directly carries through to an improved constant in \cref{thm:main,thm:distances-known}.

In general, the constants in our upper and lower bounds do not match. Closing the gaps for all of the notions of representation studied here would be of interest, both in the model with known metric space and with ordinal information. Another possible direction of interest is to understand under what natural conditions about the metric space a constant factor in representativeness can be achieved without resource augmentation.

\subsubsection*{Acknowledgements}
We would like to thank Evi Micha for useful discussions, and Jannik Peters for pointing us to additional related and simultaneous work.
This work was supported under ARO MURI grant W911NF1810208.

\bibliographystyle{plainnat}
\bibliography{../davids-bibliography/names,../davids-bibliography/conferences,../davids-bibliography/publications,../davids-bibliography/bibliography,../ref}

\newpage

\appendix

\section{Lower-Bound Examples}
\label{sec:limits}

\subsection{Lower Bound for large $\alpha$}
\label{ex:away-from-one}

In this section, we prove \cref{prop:bounded-from-optimal}, showing that for every $\alpha > 1$, there are instances $(V,c,d,k)$ for which the $(\alpha, 1+1/(2\alpha))$-core is empty.
  
\begin{proof}[Proof of \cref{prop:bounded-from-optimal}]
Given any $\alpha > 1$, let $q = \ceil{2 \alpha}$.
The committee size is $k=2q-1$.
In our instance, the voters and candidates are the same, i.e., $V=C$, and there are $n=2k$ of them; thus, the Hare Quota is $p = 2$.
The voters/candidates form two clusters $B_1, B_2$, each of size $k$.
Within each cluster, each pair has distance 1, and between clusters, the distance is large.

Let $R$ be any committee of size $k$.
For one of the two clusters (w.l.o.g.~$B_1$), the committee $R$ selects at most $q - 1$ candidates.
Let $S$ be a coalition of $q$ voters from $B_1 \setminus R$; notice that such a coalition exists, because $B_1$ contains $2q-1$ voters, of whom at most $q-1$ are in $R$.
Because $q \geq 2 \alpha$, the coalition $S$ is of size at least $\alpha \cdot p$.

Because each $v \in S$ is at distance at least 1 from its closest representative in $R$, the total cost of $S$ is at least $q$.
By including any one member of $S$ instead, the cost for that member would be reduced to 0, so the total cost would be $q-1$.
As a result, we obtain that the cost ratio is at least $1 + \frac{1}{q-1} > 1 + \frac{1}{2\alpha}$, so whenever $\beta \leq 1 + \frac{1}{2\alpha}$, we have that
$\sum_{v \in S} \min_{r \in R} d(v,r) > \beta \cdot \dsum(S, c) $ for some $c \in C \setminus R$.
In particular, $R$ cannot be in the $(\alpha, 1+1/(2\alpha))$-core, and because the argument applies to arbitrary $R$, we have shown that the $(\alpha, 1+1/(2\alpha))$-core is empty.
\end{proof}

\subsection{Lower Bound for small $\alpha$}
\label{ex:resource-augmentation}
\label{ex:info-hardness}

  In this section, we prove \cref{prop:diverging}, showing that for $\alpha$ arbitrarily close to 1, there are instances for which the $(\alpha, 1/(4(\alpha-1)))$-core is empty.

\begin{proof}[Proof of \cref{prop:diverging}]
  Consider $\alpha \in (1, 3/2)$ such that $\frac{1}{\alpha-1}$ is an integer.
  Let the committee size be $k=\frac{1}{\alpha-1}$, and consider an instance consisting of $k+1$ clusters $B_i$, each containing one candidate and $k-1$ voters.
  Thus, the total number of voters is $n=k^2-1$, and the Hare Quota is $p = \ceil{n/k} = k$.

  The distances between all voters and the candidate within a cluster are 0, whereas the distances between any pair of candidate/voter from different clusters is 1.\footnote{These distances can be implemented in a $(k+1)$-dimensional Euclidean space, by embedding cluster $i$ at distance $1/\sqrt{2}$ on the \Kth{i} coordinate axis; though this is not essential, as we allow abstract metric spaces as well. Also, notice that we could easily modify the example to be a metric instead of pseudo-metric, by replacing the distances of 0 with some small $\epsilon$.} 

  Any committee $R$ of size $k$ must omit the candidate in at least one of the clusters $B_i$.
  Consider the coalition $S$ comprising all of the $k-1$ voters in $B_i$, plus $\alpha \cdot p - (k-1) = (k+1) - (k-1) = 2$ voters from some other $B_j$.

  Because the candidate from $B_i$ is not in the committee, each of the voters in $B_i$ incurs cost at least 1, so the total cost of $B_i$ is at least $k-1$.
  On the other hand, the candidate from $B_i$ would have been an alternative leading to cost at most $2$, since the voters in $B_i$ would incur cost 0, while the voters in $B_j$ would incur cost 1.
  Thus, $\sum_{v \in S} \min_{r \in R} d(v,r) > \beta \cdot \dsum(S, c) $ for some $c \in C \setminus R$, whenever $\beta < \frac{k-1}{2} = \frac{2-\alpha}{2(\alpha-1)}$.
  Again, this implies that $R$ cannot be in the approximate core, and it applies to arbitrary $R$.
  
  Because $\alpha < 3/2$, we get that $\frac{2-\alpha}{2(\alpha-1)} > \frac{1}{4(\alpha-1)}$, implying that the $(\alpha,1/(4(\alpha-1)))$-core is empty.
\end{proof}

By modifying the proof of \cref{prop:diverging} slightly, we also obtain a slightly more refined bound on when the $(1,\beta)$-core is empty compared to \citet{LLSWW:core-fairness}, encompassing the whole range of $k$.
Specifically, we will show that for every $n,k$ and $\beta < \frac{1}{16} \cdot \min(k, n/k)$, there are instances for which the $(1,\beta)$-core is empty.
For $k=\sqrt{n}$, this recovers (up to small constants, lost only for mathematical convenience) Theorem~7 of \citet{LLSWW:core-fairness}, but extends it to arbitrary values of $k$.

Take $n,k$ as given.
If $k > n/4$, then the claimed lower bound holds trivially (because the given bound on $\beta$ is smaller than 1), so we assume $k \leq n/4$.
Again, we consider an instance with $k+1$ clusters $B_i$, each containing one candidate.
The number of voters in cluster $i$ is $b_i \in \set{\floor{n/(k+1)}, \ceil{n/(k+1)}}$, with the $b_i$ chosen such that $\sum_{i=1}^{k+1} b_i = n$.
As in the proof of \cref{prop:diverging}, distances within clusters are 0, and between clusters, the distances are 1.
The Hare Quota is $p = \ceil{n/k}$.

Again, any committee $R$ of $k$ candidates must omit at least one cluster $B_i$, and we consider the coalition $S$ of the $b_i \geq \floor{n/(k+1)}$ voters in $B_i$, plus $p-b_i$ voters from some other cluster $B_j$.
The total cost of $S$ for the committee $R$ is at least $b_i$, while it would be at most $p-b_i$ if the candidate from $B_i$ had been included. 
Thus, $\sum_{v \in S} \min_{r \in R} d(v,r) > \beta \cdot \dsum(S, c) $ for any $c \in B_i$ and $\beta < \frac{b_i}{p-b_i}$.
We now lower-bound this ratio:

  \begin{align*}
    \frac{b_i}{p-b_i}
    & \geq \frac{\floor{n/(k+1)}}{\ceil{n/k}-\floor{n/(k+1)}}
    \\ & \geq \frac{n/(k+1)-1}{n/k+2-n/(k+1)}
    \\ & = \frac{n - (k+1)}{n/k + 2(k+1)}
    \\ & \stackrel{k \leq n/4}{\geq} \frac{1}{2} \cdot \frac{n}{n/k + 2(k+1)}
    \\ & \geq \frac{1}{4} \cdot \min \left( \frac{n}{n/k}, \frac{n}{2(k+1)} \right)
    \\ & > \frac{1}{16} \cdot \min(k, n/k).
  \end{align*}

Thus, we have shown that for the given instance, the $(1, (1/16) \cdot \min(k, n/k))$-core is empty.

\subsection{Separation between Cardinal and Ordinal Proportionally Fair Clustering}
\label{ex:ordinal-vs-cardinal}

In \cref{cor:pfc}, we showed the existence of a deterministic 5.71-proportionally fair clustering algorithm under ordinal information.
On the other hand, \citet{pfc,pfc-revisited} gave a deterministic $1+\sqrt{2} \approx 2.41$-proportionally fair clustering algorithm when the metric space is known to the algorithm.
Here, we show a separation between proportionally fair clustering in the ordinal and cardinal models, by giving an instance under ordinal information for which no deterministic algorithm can be better than $2+\sqrt{5} \approx 4.23$-proportionally fair.

We consider a committee selection problem with $n=6$ voters, $m=6$ candidates, and committee size $k=3$.
The voters are $V = \set{v_1, v_2, v_3, v'_1, v'_2, v'_3}$, and the candidates are $C = \set{c_1, c_2, c_3, c'_1, c'_2, c'_3}$. 
The rankings of the voters are as follows:
  \begin{align*}
    v_1:  & c_1 \succ c_2 \succ c_3 \succ c'_1 \succ c'_2 \succ c'_3 & v'_1: & c'_1 \succ c'_2 \succ c'_3 \succ c_1 \succ c_2 \succ c_3
\\   v_2:  & c_2 \succ c_3 \succ c_1 \succ c'_1 \succ c'_2 \succ c'_3 & v'_2:  & c'_2 \succ c'_3 \succ c'_1 \succ c_1 \succ c_2 \succ c_3.
\\   v_3:  & c_3 \succ c_1 \succ c_2 \succ c'_1 \succ c'_2 \succ c'_3 & v'_3:  & c'_3 \succ c'_1 \succ c'_2 \succ c_1 \succ c_2 \succ c_3
\end{align*}

Because the committee size is $k=3$, either at most one candidate is chosen from $\set{c_1, c_2, c_3}$ or at most one candidate is chosen from $\set{c'_1, c'_2, c'_3}$.
Without loss of generality, assume that at most one candidate is chosen from $\set{c_1, c_2, c_3}$.

The metric space we define will have two clusters, one comprising $c_1, c_2, c_3, v_1, v_2, v_3$, and the other $c'_1, c'_2, c'_3, v'_1, v'_2, v'_3$.
The two clusters are far from each other.

If no candidate from $\set{c_1, c_2, c_3}$ is chosen, then the voter coalition $S = \set{v_1, v_2}$ will have extremely high cost (the distance between the clusters), and strongly prefer to deviate to any candidate in $\set{c_1, c_2, c_3}$. 
So for the rest of the analysis, we assume that exactly one candidate from $\set{c_1, c_2, c_3}$ is chosen.
We assume that the chosen candidate is $c_1$ --- the construction below can be altered straightforwardly simply by switching the names of candidates if another candidate were chosen.

Consider the following distances between the voters and candidates, with $\delta = \frac{\sqrt{5}-1}{2}$, and $\epsilon$ some arbitrarily small number (used solely to avoid ties):

\begin{center}
  \begin{tabular}{c|c|c|c}
      &  $c_1$ &  $c_2$            &  $c_3$ \\ \hline
  $ v_1 (c_1 \succ c_2 \succ c_3)$ &   3  &  $3+\epsilon$   &  $3+2\epsilon$ \\ \hline
  $ v_2 (c_2 \succ c_3 \succ c_1)$ &   $3+2\delta$  &  1              &  $1+\epsilon$ \\ \hline
  $ v_3 (c_3 \succ c_1 \succ c_2)$ &   $2+\delta$  &  $2+\delta+\epsilon$   &  $\delta$
  \end{tabular}
\end{center}

The distances between $c'_1,c'_2,c'_3$ and $v'_1, v'_2, v'_3$ are all 1, and the distances between the two clusters are 100.
 It can be verified that all of these distances together satisfy the triangle inequality (the only interesting cases occur between $c_1,c_2,c_3$ and $v_2, v_3$), and are consistent with the rankings.

Now consider the coalition $\set{v_2,v_3}$.
For each of them, the best candidate in the committee is $c_1$, with respective costs $3+2\delta$ and $2+\delta$.
By deviating to $c_3$, their new respective costs would be $1+\epsilon$ and $\delta$.
The resulting cost improvement would be $\min(\frac{3+2\delta}{1+\epsilon}, \frac{2+\delta}{\delta}) \geq 3+2\delta - O(\epsilon)$.
Thus, as $\epsilon \to 0$, we see that no proportional fairness guarantee of $\gamma < 3 + 2\delta = \sqrt{5}+2 \approx 4.236$ can be possible information-theoretically.

\end{document}